\documentstyle[prd,preprint,tighten,aps,amssymb,newlfont,epsfig]{revtex}
\begin{document}
\draft
\preprint{
\begin{tabular}{r}
DFTT 68/97\\
IASSNS-AST 97/66\\
UWThPh-1997-44\\
hep-ph/9711416
\end{tabular}
}
\title{Neutrino masses and mixing from neutrino oscillation experiments}
\author{S.M. Bilenky}
\address{Joint Institute for Nuclear Research, Dubna, Russia, and\\
Institute for Advanced Study, Princeton, N.J. 08540}
\author{C. Giunti}
\address{INFN, Sezione di Torino, and Dipartimento di Fisica Teorica,
Universit\`a di Torino,\\
Via P. Giuria 1, I--10125 Torino, Italy}
\author{W. Grimus}
\address{Institute for Theoretical Physics, University of Vienna,\\
Boltzmanngasse 5, A--1090 Vienna, Austria}
\maketitle
\begin{abstract}
We discuss which information
on neutrino masses and mixing can be obtained
from the results of all neutrino oscillation experiments
in the cases of three and four massive neutrinos.
We show that in the three-neutrino case the neutrino oscillation data
are not compatible with a hierarchy of couplings.
In the case of four neutrinos,
a hierarchy of masses
is disfavored by the data and
only two schemes with two pairs of neutrinos with close masses
separated by a gap of the order of 1 eV
can accommodate the results of all experiments.
\end{abstract}

\pacs{Talk presented by C. Giunti at the
\emph{International Europhysics Conference on High Energy Physics},
19--26 August 1997, Jerusalem, Israel.}

The search for neutrino oscillations
is one of the most active branches of today's
high-energy physics.
From the LEP measurements
of the invisible width of the $Z$-boson
we know that there are three light active flavor neutrinos:
$\nu_e$,
$\nu_\mu$ and
$\nu_\tau$.
In general,
flavor neutrinos are not mass eigenstates
and the left-handed flavor neutrino
fields
$\nu_{{\alpha}L}$
are superpositions
of
the left-handed components
$\nu_{kL}$
of the fields of neutrinos with a definite mass
($k=1,2,3,\ldots,n$):
$
\nu_{{\alpha}L}
=
\sum_{k=1}^{n}
U_{{\alpha}k}
\,
\nu_{kL}
$,
where $U$
is a unitary mixing matrix.
The number $n$ of massive neutrinos can be three or more,
without any experimental upper limit.
If $n>3$, there are
$n-3$ sterile flavor neutrino fields,
i.e.,
fields of
neutrinos which do not take part in standard weak interactions;
in this case
$
\nu_\alpha
=
\nu_e,\nu_\mu,\nu_\tau,\nu_{s_{1}},\nu_{s_{2}},\ldots,\nu_{s_{n-3}}
$.
Neutrino oscillations
is a direct consequence of neutrino mixing;
the probability of
$\nu_\alpha\to\nu_\beta$
transitions
is given by
(see \cite{BP78})
\begin{equation}
P_{\nu_\alpha\to\nu_\beta}
=
\left|
\sum_{k=1}^{n}
U_{{\beta}k}
U_{{\alpha}k}^{*}
\exp\left( - i \frac{ \Delta{m}^2_{k1} L }{ 2 E } \right)
\right|^2
,
\label{151}
\end{equation}
where
$L$ is the
distance between
the neutrino source and detector,
$E$ is the neutrino energy
and
$ \Delta{m}^{2}_{kj} \equiv m_k^2 - m_j^2 $.

In this report we discuss which information
on the neutrino mass spectrum and mixing parameters
can be obtained from the results of
neutrino oscillation experiments.
Many short-baseline (SBL)
neutrino oscillation experiments
with reactor and accelerator neutrinos
did not find any evidence
of neutrino oscillations.
Their results can be used in order to constrain
the allowed values of the neutrino masses
and of the elements of the mixing matrix.
In our analysis
we use the most stringent exclusion plots
obtained in the
$\bar\nu_e\to\bar\nu_e$
channel
by the Bugey experiment \cite{Bugey95},
in the
$\nu_\mu\to\nu_\mu$
channel
by the CDHS and CCFR experiments \cite{CDHS84-CCFR84},
and in the
$
\stackrel{\makebox[0pt][l]
{$\hskip-3pt\scriptscriptstyle(-)$}}{\nu_{\mu}}
\to
\stackrel{\makebox[0pt][l]
{$\hskip-3pt\scriptscriptstyle(-)$}}{\nu_{e}}
$
channel
by the
BNL E734,
BNL E776
and
CCFR
\cite{BNLE734-BNLE776-CCFR96}
experiments.

There are three experimental indications
in favor of neutrino oscillations that come from the anomalies observed by
the solar neutrino experiments \cite{sunexp},
the atmospheric neutrino experiments \cite{atmexp}
and the LSND experiment \cite{LSND}.
The solar neutrino deficit can be explained
with oscillations of solar $\nu_e$'s
into other states
and indicates a mass-squared difference
of the order of
$ 10^{-5} \, \mathrm{eV}^2 $
in the case of resonant MSW transitions
or
$ 10^{-10} \, \mathrm{eV}^2 $
in the case of vacuum oscillations.
The atmospheric neutrino anomaly
can be explained by
$
\stackrel{\makebox[0pt][l]
{$\hskip-3pt\scriptscriptstyle(-)$}}{\nu_{\mu}}
\to
\stackrel{\makebox[0pt][l]
{$\hskip-3pt\scriptscriptstyle(-)$}}{\nu_{x}}
$
oscillations ($x\neq\mu$)
with 
a mass-squared difference
of the order of
$ 10^{-2} \, \mathrm{eV}^2 $.
Finally,
the LSND experiment found
indications in favor of
$ \bar\nu_\mu \to \bar\nu_e $
oscillations
with a mass-squared difference
of the order of
$ 1 \, \mathrm{eV}^2 $.

Hence,
three different scales of mass-squared difference
are needed in order to explain the three indications
in favor of neutrino oscillations.
This means that the number of massive neutrinos
must be bigger than three.
In the following we consider the simplest possibility,
i.e.
the existence of four massive neutrinos ($n=4$).
In this case,
besides the three light flavor neutrinos
$\nu_e$,
$\nu_\mu$,
$\nu_\tau$,
there is a light sterile neutrino
$\nu_s$.

However,
before considering the case of four neutrinos,
we discuss the minimal possibility
of the existence of only three massive neutrinos ($n=3$).
In this case
one of the experimental anomalies mentioned above
cannot be explained with neutrino oscillations
(we choose to disregard the atmospheric neutrino anomaly).

In both cases of three and four massive neutrinos
the oscillations in the LSND experiment
imply that
the largest mass-squared difference
$ \Delta{m}^{2}_{n1} \equiv m^2_n - m^2_1 $
is relevant for SBL oscillations,
whereas the other mass-squared differences are much smaller.
Hence, the mass spectrum must be composed of
two groups of massive neutrinos
with close masses
($ \nu_1, \ldots , \nu_{r-1} $
and
$ \nu_{r}, \ldots , \nu_n $)
separated by a mass difference in the eV range
($
m_1 < \ldots < m_{r-1}
\ll
m_r < \ldots < m_n
$)
and
in SBL experiments we have
$ \frac{ \Delta{m}^{2}_{n1} L }{ 2 E } \gtrsim 1 $,
$ \frac{ \Delta{m}^{2}_{k1} L }{ 2 E } \ll 1 $
for
$  k < r $,
$ \frac{ \Delta{m}^{2}_{nk} L }{ 2 E } \ll 1 $
for
$ k \geq r $.
The formula (\ref{151})
written as
\begin{equation}
P_{\nu_\alpha\to\nu_\beta}
=
\left|
\sum_{k=1}^{r-1}
U_{{\beta}k}
U_{{\alpha}k}^{*}
e^{ - i \frac{ \Delta{m}^{2}_{k1} L }{ 2 E } }
+
e^{ - i \frac{ \Delta{m}^{2}_{n1} L }{ 2 E } }
\sum_{k=r}^{n}
U_{{\beta}k}
U_{{\alpha}k}^{*}
e^{ i \frac{ \Delta{m}^{2}_{nk} L }{ 2 E } }
\right|^2
\label{152}
\end{equation}
leads to the following expression
for the transition ($\beta\neq\alpha$)
and survival ($\beta=\alpha$) probabilities
of neutrinos and anti-neutrinos
in SBL experiments:
\begin{equation}
P^{(\mathrm{SBL})}_{\stackrel{\makebox[0pt][l]
{$\hskip-3pt\scriptscriptstyle(-)$}}{\nu_{\alpha}}
\to\stackrel{\makebox[0pt][l]
{$\hskip-3pt\scriptscriptstyle(-)$}}{\nu_{\beta}}}
=
A_{\alpha;\beta}
\,
\sin^2 \frac{ \Delta{m}^{2} L }{ 4 E }
,
\qquad
P^{(\mathrm{SBL})}_{\stackrel{\makebox[0pt][l]
{$\hskip-3pt\scriptscriptstyle(-)$}}{\nu_{\alpha}}
\to\stackrel{\makebox[0pt][l]
{$\hskip-3pt\scriptscriptstyle(-)$}}{\nu_{\alpha}}}
=
1
-
B_{\alpha;\alpha}
\,
\sin^2 \frac{ \Delta{m}^{2} L }{ 4 E }
,
\label{108}
\end{equation}
with
$ \Delta{m}^{2} \equiv \Delta{m}^{2}_{n1} $
and
the oscillation amplitudes
\begin{equation}
A_{\alpha;\beta}
=
4 \left| \sum_{k=r}^{n} U_{{\beta}k} \, U_{{\alpha}k}^{*} \right|^2
,
\quad
B_{\alpha;\alpha}
=
4
\left( \sum_{k=r}^{n} |U_{{\alpha}k}|^2 \right)
\left( 1 - \sum_{k=r}^{n} |U_{{\alpha}k}|^2 \right)
.
\label{111}
\end{equation}
The formulas (\ref{108})
have the same form of the standard expressions
for the oscillation probabilities in the case of
two neutrinos
(see \cite{BP78})
with which
the data of all the SBL experiments
have been analyzed by the experimental groups.
Hence,
the results of these analyses
can be used
in order to constrain
the possible values of
the oscillation amplitudes
$A_{\alpha;\beta}$
and
$B_{\alpha;\alpha}$.

First,
we consider the scheme 3H of Tab.\ref{tab1},
with three neutrinos
and a mass hierarchy.
This scheme
(as all the schemes with three neutrinos)
provides only two independent
mass-squared differences,
$\Delta{m}^{2}_{21}$
and
$\Delta{m}^{2}_{31}$,
which
we choose to be relevant for the solution of the
solar neutrino problem
and
for neutrino oscillations in the LSND experiment,
respectively.

Let us emphasize that the mass spectrum 3H
with three neutrinos and a mass hierarchy
is the simplest and most natural one,
being analogous to the mass spectra of
charged leptons, up and down quarks.
Moreover,
a scheme with three neutrinos and a mass hierarchy
is predicted by the see-saw mechanism
for the generation of neutrino masses,
which can explain the smallness of the neutrino masses
with respect to the masses of the corresponding
charged leptons.

In the case of scheme 3H we have
$n=r=3$ and
Eq.(\ref{111})
implies that
\begin{equation}
A_{\alpha;\beta}
=
4
\,
|U_{{\alpha}3}|^2
\,
|U_{{\beta}3}|^2
\,,
\qquad
B_{\alpha;\alpha}
=
4
\,
|U_{{\alpha}3}|^2
\left(
1
-
|U_{{\alpha}3}|^2
\right)
\,.
\label{AB3}
\end{equation}
Hence,
neutrino oscillations in
SBL experiments
depend on three parameters:
$\Delta{m}^{2}\equiv\Delta{m}^{2}_{31}$,
$|U_{e3}|^2$ and
$|U_{\mu3}|^2$
(the unitarity of $U$ implies that
$
|U_{\tau3}|^2
=
1
-
|U_{e3}|^2
-
|U_{\mu3}|^2
$).
From the exclusion plots obtained in
reactor $\bar\nu_e$
and accelerator
$\nu_\mu$
disappearance experiments
it follows that
at any fixed value of $\Delta{m}^2$,
the oscillation amplitudes
$B_{e;e}$
and
$B_{\mu;\mu}$
are bounded by the upper values
$B_{e;e}^{0}$
and
$B_{\mu;\mu}^{0}$,
respectively,
which are small quantities for
$
0.3
\lesssim
\Delta{m}^2
\lesssim
10^{3} \, \mathrm{eV}^2
$.
From Eq.(\ref{AB3})
one can see that small upper bounds for
$B_{e;e}$
and
$B_{\mu;\mu}$
imply that the parameters
$|U_{e3}|^2$
and
$|U_{\mu3}|^2$
can be either small or large (i.e., close to one):
\begin{equation}
|U_{{\alpha}3}|^2
\leq
a^0_\alpha
\quad \mbox{or} \quad
|U_{{\alpha}3}|^2
\geq
1 - a^0_\alpha
\,,
\quad \mbox{with} \quad
a^0_\alpha
=
\frac{1}{2}
\left(
1
-
\sqrt{ 1 - B_{\alpha;\alpha}^{0} }
\right)
,
\label{a0}
\end{equation}
for $\alpha=e,\mu$.
Both
$a^{0}_{e}$
and
$a^{0}_{\mu}$
are small
($ a^{0}_e \lesssim 4 \times 10^{-2} $
and
$ a^{0}_\mu \lesssim 2 \times 10^{-1} $)
for any value of
$\Delta{m}^{2}$
in the range
$
0.3
\lesssim
\Delta{m}^2
\lesssim
10^{3} \, \mathrm{eV}^2
$
(see Fig.1 of Ref.\cite{BBGK96}).

Since large values of both
$|U_{e3}|^2$
and
$|U_{\mu3}|^2$
are excluded by the unitarity of the mixing matrix
($|U_{e3}|^2+|U_{\mu3}|^2\leq1$),
at any fixed value of $\Delta{m}^2$
there are three regions in the
$|U_{e3}|^2$--$|U_{\mu3}|^2$
plane
which are allowed by
the exclusion plots of SBL disappearance experiments:
Region I,
with
$ |U_{e3}|^2 \leq a^{0}_{e} $
and
$ |U_{\mu3}|^2 \leq a^{0}_{\mu} $;
Region II,
with
$ |U_{e3}|^2 \leq a^{0}_{e} $
and
$ |U_{\mu3}|^2 \geq 1 - a^{0}_{\mu} $;
Region III,
with
$ |U_{e3}|^2 \geq 1 - a^{0}_{e} $
and
$ |U_{\mu3}|^2 \leq a^{0}_{\mu} $.

In region III
$ |U_{e3}|^2 $
is large and $\nu_e$
has a small mixing with $\nu_1$ and $\nu_2$,
which is insufficient for the explanation
of the solar neutrino problem.
Indeed,
the survival probability of solar $\nu_e$'s
is bounded by
$
P_{\nu_e\to\nu_e}^{\mathrm{sun}}
\geq
|U_{e3}|^4
$
(see \cite{BBGK96}).
If 
$ |U_{e3}|^2 \geq 1 - a^0_e $,
we have
$
P_{\nu_e\to\nu_e}^{\mathrm{sun}}
\gtrsim
0.92
$
at all neutrino energies,
which is a bound that is not compatible
with the solar neutrino data.
Hence,
region III is excluded by solar neutrinos.

In region I
$ A_{\mu;e} \leq 4 \, a^0_e \, a^0_\mu $,
which means that the probability of
$
\stackrel{\makebox[0pt][l]
{$\hskip-3pt\scriptscriptstyle(-)$}}{\nu_{\mu}}
\leftrightarrows
\stackrel{\makebox[0pt][l]
{$\hskip-3pt\scriptscriptstyle(-)$}}{\nu_{e}}
$
transitions in SBL experiments is strongly suppressed.
The corresponding upper bound obtained
from the 90\% CL exclusion plots of the
Bugey \cite{Bugey95}
$\bar\nu_e$
disappearance experiment
and of the
CDHS and CCFR \cite{CDHS84-CCFR84}
$\nu_\mu$
disappearance experiments
is represented in Fig.\ref{fig1}
by the curve passing trough the circles.
The shadowed regions in Fig.\ref{fig1}
are allowed at 90\% CL by the results of the LSND experiment.
Also shown
are the 90\% CL exclusion curves found in the
BNL E734,
BNL E776
and
CCFR
\cite{BNLE734-BNLE776-CCFR96}
$
\stackrel{\makebox[0pt][l]
{$\hskip-3pt\scriptscriptstyle(-)$}}{\nu_{\mu}}
\to
\stackrel{\makebox[0pt][l]
{$\hskip-3pt\scriptscriptstyle(-)$}}{\nu_{e}}
$
appearance experiments
and in the Bugey experiment.
One can see from Fig.\ref{fig1} that
in region I,
the bounds obtained from the results of
$\bar\nu_e\to\bar\nu_e$,
$\nu_\mu\to\nu_\mu$
and
$
\stackrel{\makebox[0pt][l]
{$\hskip-3pt\scriptscriptstyle(-)$}}{\nu_{\mu}}
\to
\stackrel{\makebox[0pt][l]
{$\hskip-3pt\scriptscriptstyle(-)$}}{\nu_{e}}
$
experiments
are not compatible 
with the allowed regions of the LSND experiment
\cite{BBGK96}.
Therefore,
region I
is disfavored by the results
of SBL experiments.
This is an important indication,
because
region I
is the only one in which it is possible to have
a hierarchy of
the elements of the neutrino mixing matrix
analogous to the one of the
quark mixing matrix.

Having excluded the regions I and III
of the scheme 3H,
we are left only with region II,
where $\nu_\mu$ has a large mixing
with $\nu_3$,
i.e.,
$\nu_\mu$
(not $\nu_\tau$)
is the ``heaviest'' neutrino.

Let us now consider the possible schemes with four neutrinos,
which provide three independent
mass-squared differences
and allow to accommodate
in a natural way
all the three experimental indications
in favor of neutrino oscillations.
We consider first the scheme
4H of Tab.\ref{tab1}
with four neutrinos and
a mass hierarchy.
The three independent
mass-squared differences,
$\Delta{m}^{2}_{21}$,
$\Delta{m}^{2}_{31}$ and
$\Delta{m}^{2}_{41}$,
are taken to be relevant for the oscillations
of solar, atmospheric and LSND neutrinos,
respectively.
In the case of scheme 4H
we have
$n=r=4$
and Eq.(\ref{111})
implies that
the oscillation amplitudes are given by
\begin{equation}
A_{\alpha;\beta}
=
4
\,
|U_{{\alpha}4}|^2
\,
|U_{{\beta}4}|^2
\,,
\qquad
B_{\alpha;\alpha}
=
4
\,
|U_{{\alpha}4}|^2
\left(
1
-
|U_{{\alpha}4}|^2
\right)
\,.
\label{AB4}
\end{equation}
In this case,
neutrino oscillations in
SBL experiments
depend on four parameters:
$\Delta{m}^{2}\equiv\Delta{m}^{2}_{31}$,
$|U_{e3}|^2$,
$|U_{\mu3}|^2$ and
$|U_{\tau3}|^2$.
From the similarity of the amplitudes
(\ref{AB4})
with the corresponding ones
given in Eq.(\ref{AB3}),
it is clear that
replacing
$|U_{{\alpha}3}|^2$
with
$|U_{{\alpha}4}|^2$
we can apply
to the scheme 4H
the same analysis presented for the scheme 3H.
Hence,
also the regions III and I of the scheme 4H are excluded,
respectively,
by the solar neutrino problem
and by the results of SBL experiments.
Furthermore,
the purpose of considering the scheme 4H
is to have the possibility to explain the atmospheric neutrino anomaly,
but this is not possible if the
neutrino mixing parameters lie in region II.
Indeed,
in region II
$ |U_{\mu4}|^2 $
is large
and the muon neutrino has
a small mixing with the light neutrinos
$\nu_1$,
$\nu_2$ and
$\nu_3$,
which is insufficient for the explanation of
the atmospheric neutrino anomaly
\cite{BGG96}.

Hence,
the scheme 4H
is disfavored by the results of
neutrino oscillation experiments.
For the same reasons,
all possible schemes with four neutrinos
and a mass spectrum in which
three masses are clustered and one mass is separated from the
others by a gap of about 1 eV
(needed for the explanation of the LSND data)
are disfavored by the results
of neutrino oscillation experiments.
Therefore,
there are only two possible schemes
with four neutrinos
which are compatible with the results
of all neutrino oscillation experiments:
the schemes 4A and 4B of Tab.\ref{tab1}.
In these two schemes
the four neutrino masses
are divided in two pairs of close masses
separated by a gap of about 1 eV.
In scheme A,
$\Delta{m}^{2}_{21}$
is relevant
for the explanation of the atmospheric neutrino anomaly
and
$\Delta{m}^{2}_{43}$
is relevant
for the suppression of solar $\nu_e$'s.
In scheme B,
the roles of
$\Delta{m}^{2}_{21}$
and
$\Delta{m}^{2}_{43}$
are reversed.

From Eq.(\ref{111})
and using the unitarity of the mixing matrix,
the oscillation amplitudes
$B_{\alpha;\alpha}$
in the schemes 4A and 4B
($n=4$, $r=3$)
are given by
$
B_{\alpha;\alpha}
=
4 \, c_{\alpha} \left( 1 - c_{\alpha} \right)
$,
with
$
c_{\alpha}
\equiv
\sum_{k=1,2}
|U_{{\alpha}k}|^2
$
in the scheme 4A
and
$
c_{\alpha}
\equiv
\sum_{k=3,4}
|U_{{\alpha}k}|^2
$
in the scheme 4B.
This expression for
$B_{\alpha;\alpha}$
has the same form as the one in Eq.(\ref{AB4}),
with
$|U_{{\alpha}4}|^2$
replaced by $c_{\alpha}$.
Therefore,
we can apply the same analysis presented for the scheme 4H
and we obtain four allowed regions
in the
$c_e$--$c_\mu$
plane
(now the region with large
$c_e$ and $c_\mu$
is not excluded by
the unitarity of the mixing matrix,
which gives the constraint
$ c_e + c_\mu \leq 2 $):
Region I,
with
$ c_e \leq a^{0}_{e} $
and
$ c_\mu \leq a^{0}_{\mu} $;
Region II,
with
$ c_e \leq a^{0}_{e} $
and
$ c_\mu \geq 1 - a^{0}_{\mu} $;
Region III,
with
$ c_e \geq 1 - a^{0}_{e} $
and
$ c_\mu \leq a^{0}_{\mu} $;
Region IV,
with
$ c_e \geq 1 - a^{0}_{e} $
and
$ c_\mu \geq 1 - a^{0}_{\mu} $.
Following the same reasoning
as in the case of scheme 4H,
one can see that
the regions III and IV
are excluded by the solar neutrino data
and the regions I and III
are excluded by the results of the atmospheric neutrino experiments
\cite{BGG96}.
Hence,
only region II
is allowed by the results of all experiments.

If the neutrino mixing parameters lie in region II,
in the scheme 4A (4B) the electron (muon)
neutrino is ``heavy'',
because it has a large mixing with $\nu_3$ and $\nu_4$,
and the muon (electron) neutrino is ``light''.
Thus,
the schemes 4A and 4B give different predictions
for the effective Majorana mass
$
\langle{m}\rangle
=
\sum_{k} U_{ek}^2 m_k
$
in
neutrinoless double-beta decay
experiments:
since
$ m_3 \simeq m_4 \gg m_1 \simeq m_2 $,
we have
$
|\langle{m}\rangle|
\leq
(1-c_{e}) m_4
\simeq
m_4
$
in the scheme 4A
and
$
|\langle{m}\rangle|
\leq
c_e m_4
\leq
a^{0}_{e} m_4
\ll m_4
$
in the scheme 4B.
Hence,
if the scheme 4A is realized in nature,
the experiments on the search for neutrinoless double-beta
decay can reveal
the effects of the heavy
neutrino masses $ m_3 \simeq m_4 $.
Furthermore,
the smallness of $c_e$
in both schemes 4A and 4B
implies that the electron neutrino has a
small mixing with the neutrinos whose mass-squared difference is
responsible for the oscillations of atmospheric neutrinos
(i.e.,
$\nu_1$, $\nu_2$ in scheme 4A and $\nu_3$, $\nu_4$ in scheme
4B).
Hence,
the transition probability of
electron neutrinos and antineutrinos
into other states
in atmospheric and long-baseline experiments
is suppressed
\cite{BGG97}\footnote{After
we finished this paper the results of the first
long-baseline reactor experiment CHOOZ appeared
(M. Apollonio \emph{et al.}, preprint hep-ex/9711002).
No indications in favor of
$\bar\nu_e\to\bar\nu_e$
transitions were found in this experiment.
The upper bound for the
transition probability
of electron antineutrinos into other states
found in the CHOOZ experiment is in agreement with
the limit obtained in Ref.\cite{BGG97}.}.

\newpage

\begin{minipage}[p]{0.95\textwidth}
\begin{center}
\begin{tabular}{rl}
3H:
\quad
&
$
\underbrace{
\overbrace{m_1 \ll m_2}^{\mathrm{sun}}
\ll
m_3
}_{\mathrm{LSND}}
$
\\ & \\
4H:
\quad
&
$
\underbrace{
\overbrace{
\overbrace{m_1 \ll m_2}^{\mathrm{sun}}
\ll m_3
}^{\mathrm{atm}}
\ll m_4
}_{\mathrm{LSND}}
$
\\ & \\
4A:
\quad
&
$
\underbrace{
\overbrace{m_1 < m_2}^{\mathrm{atm}}
\ll
\overbrace{m_3 < m_4}^{\mathrm{sun}}
}_{\mathrm{LSND}}
$
\\ & \\
4B:
\quad
&
$
\underbrace{
\overbrace{m_1 < m_2}^{\mathrm{sun}}
\ll
\overbrace{m_3 < m_4}^{\mathrm{atm}}
}_{\mathrm{LSND}}
$
\end{tabular}
\end{center}
\end{minipage}
\begin{center}
\vspace{1cm}
\refstepcounter{table}
\label{tab1}
\Large Table \ref{tab1}
\end{center}

\newpage

\begin{minipage}[p]{0.95\textwidth}
\begin{center}
\mbox{\epsfig{file=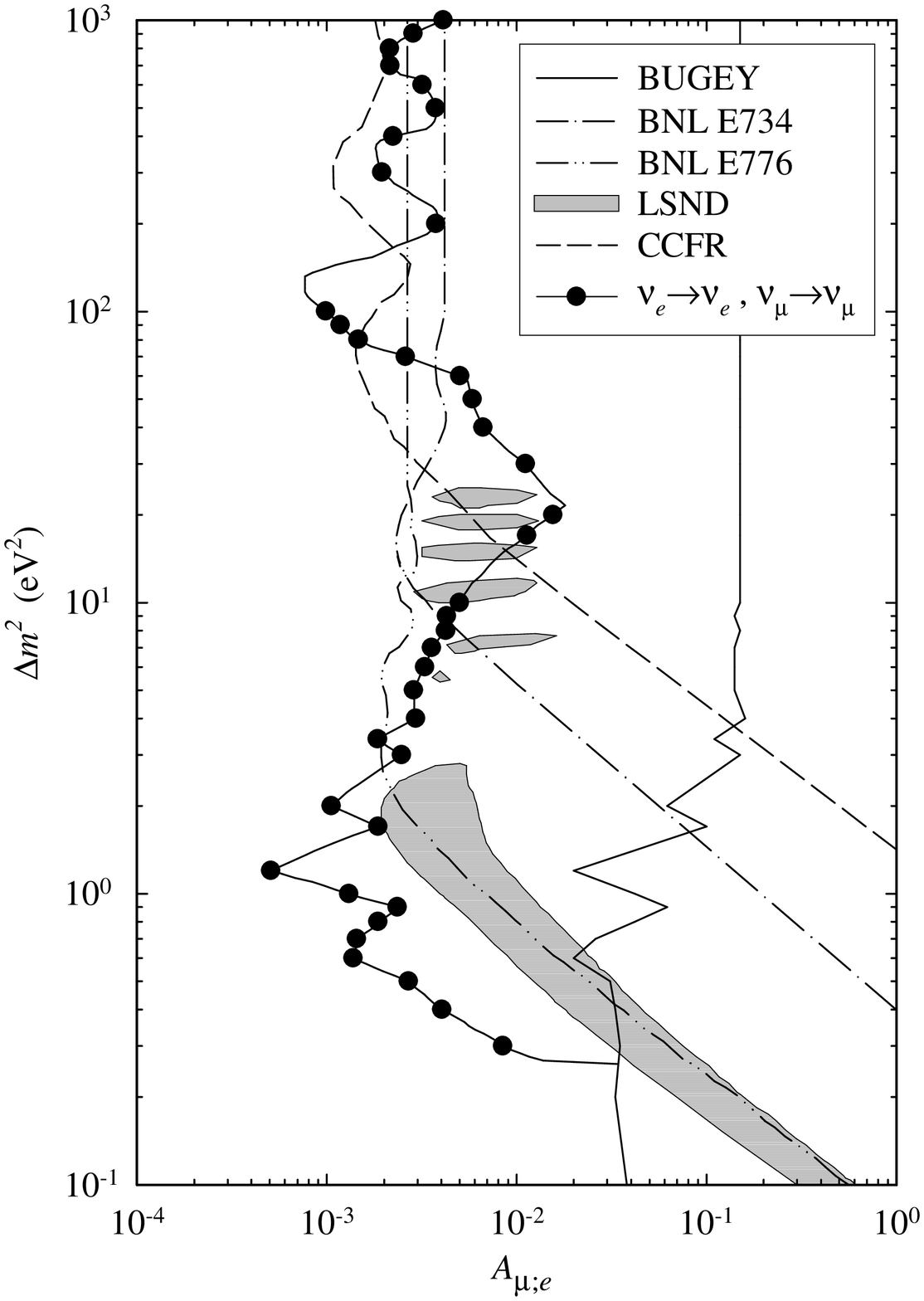,height=0.95\textheight}}
\end{center}
\end{minipage}
\begin{center}
\refstepcounter{figure}
\label{fig1}                 
\Large Figure \ref{fig1}
\end{center}

\end{document}